\documentclass[12pt]{article}
 \usepackage{epsfig}
 \def\be{\begin{equation}}
 \def\ee{\end{equation}}
 \def\bea{\begin{eqnarray}}
 \def\eea{\end{eqnarray}}
 \usepackage{graphicx}

 \catcode`\@=11
 \def\lsim{\mathrel{\mathpalette\@versim<}}
 \def\gsim{\mathrel{\mathpalette\@versim>}}
 \def\@versim#1#2{\vcenter{\offinterlineskip
 \ialign{$\m@th#1\hfil##\hfil$\crcr#2\crcr\sim\crcr } }}
 \catcode`\@=12

 \catcode`@=12
\begin{document}
 \thispagestyle{empty}
 \begin{flushright}
 UCRHEP-T625\\
 Sep 2023\
 \end{flushright}
 \vspace{0.6in}
 \begin{center}
 {\LARGE \bf Parallel Seesaw Mechanisms for Neutrinos and\\ 
Freeze-In Long-Lived Dark Matter\\}
 \vspace{1.5in}
 {\bf Ernest Ma\\}
 \vspace{0.1in}
{\sl Department of Physics and Astronomy,\\ 
University of California, Riverside, California 92521, USA\\}
\end{center}
 \vspace{1.2in}

\begin{abstract}\
If dark matter is light, it may be due to a seesaw mechanism just as 
neutrinos are.  It is postulated that both originate from the same type of 
heavy fermion anchors, either singlets or triplets.  In the latter case, 
a shift of the $W$ mass is predicted, as suggested by the $CDF$ precision 
measurement.  A spontaneously broken dark $U(1)$ gauge symmetry is assumed, 
resulting in freeze-in long-lived light dark matter.
\end{abstract}

\newpage
\baselineskip 24pt

\noindent \underline{\it Introduction}~:~ 
Neutrinos are elusive because they are light and weakly interacting.  Dark 
matter is also elusive and yet to be observed directly.  Perhaps the reason 
is the same, namely that dark matter is light and feebly interacting.  As 
opposed to thermal freeze-out for heavy dark matter, it is produced through 
freeze-in~\cite{mmy93,ckkr01,m02,aim07,hjmw10} as the decay product of some 
other particle, such as~\cite{m19} the Higgs 
boson of the Standard Model (SM).  It may also be very long-lived and does 
not require an unbroken symmetry which would be necessary in the case of 
the usually assumed stable dark matter.

To understand why neutrinos are light, there is the well-known seesaw 
mechanism~\cite{m98}.  If dark matter is also light, the reason may well 
be the same. To explore this idea in detail, two scenarios are studied 
where the anchors of both seesaw mechanisms are assumed to be of the same 
type.  One is an extension of Type I seesaw with heavy neutrino singlets 
as the anchors, the other is its Type III analog, using heavy lepton triplets 
instead~\cite{flhj89,m09}.  In the latter case, a scalar triplet is also 
present, resulting in a shift~\cite{s22,ky22,m22,h22,goorst23,ps23} 
of the $W$ boson mass from the prediction of the SM, as indicated by recent 
$CDF$ data~\cite{cdf22}. 

\noindent \underline{\it Dark $U(1)_D$ Gauge Symmetry}~:~ 
To implement the idea of a seesaw mass for dark matter, a dark $U(1)_D$ gauge 
symmetry is postulated.  The relevant particles in the two proposed 
scenarios are listed in Tables 1 and 2.
\begin{table}[thb]
\centering
\begin{tabular}{|c|c|c|c|}
\hline
fermion/scalar & $SU(2)_L$ & $U(1)_Y$ & $U(1)_D$ \\ 
\hline
$(\nu,l)_L$ & 2 & $-1/2$ & 0 \\ 
$l_R$ & 1 & $-1$ & 0 \\ 
$N_R,N'_R$ & 1 & 0 & 0 \\ 
\hline
$\Phi = (\phi^+,\phi^0)$ & 2 & 1/2 & 0 \\ 
\hline
\hline
$S_L$ & 1 & 0 & 1 \\ 
\hline
$\chi^0$ & 1 & 0 & 1 \\ 
\hline
\end{tabular}
\caption{Seesaw dark matter with heavy neutrino singlet anchors.}
\end{table}

\noindent The model of Table 1 assumes the usual Type I seesaw mechanism 
where three heavy neutrino singlets $N_R$ act as anchors for imparting 
small masses to the three observed neutrinos $\nu_L$, through the Yukawa 
terms 
\begin{equation}
f_N \overline{N_R} (\nu_L \phi^0 - l_L \phi^+).
\end{equation}  
The neutrinos themselves 
have no invariant mass terms in the Lagrangian because they transform as 
doublets under the $SU(2)_L \times U(1)_Y$ gauge symmetry. 
In the dark sector, the $U(1)_D$ gauge symmetry is spontaneously broken 
by $\chi^0$, but $S_L$ has no invariant mass.  It acquires a small mass 
from $N'_R$ as well through the Yukawa term 
\begin{equation}
f_\chi \overline{S_L} N'_R \chi^0.
\end{equation} 
(The question of anomaly cancellation will be discussed later.) For all 
three neutrinos and $S_L$ to acquire seesaw masses, there should be three 
$N_R$ copies and one $N'_R$.  They are distinguished by the dimension-four 
Yukawa terms of Eqs.~(1) and (2), where $\nu,l,N$ may be chosen to be odd 
under lepton parity, and $S,N'$ to be odd under dark parity.  However, 
both $N$ and $N'$ do not transform under the gauge symmetries of the model, 
so they can mix.  It is assumed here that they do so only through the 
dimension-three soft $N N'$ Majorana mass terms, which will be assumed 
small in the following.  The justification is that their absence would 
enhance the symmetries of the model, an argument attributed to 't Hooft 
and often used in many such discussions.

\begin{table}[thb]
\centering
\begin{tabular}{|c|c|c|c|}
\hline
fermion/scalar & $SU(2)_L$ & $U(1)_Y$ & $U(1)_D$ \\ 
\hline
$(\nu,l)_L$ & 2 & $-1/2$ & 0 \\ 
$l_R$ & 1 & $-1$ & 0 \\ 
$(\Sigma^+,\Sigma^0,\Sigma^-)_R$ & 3 & 0 & 0 \\ 
$({\Sigma'}^+,{\Sigma'}^0,{\Sigma'}^-)_R$ & 3 & 0 & 0 \\ 
\hline
$\Phi = (\phi^+,\phi^0)$ & 2 & 1/2 & 0 \\ 
\hline
\hline
$S_L$ & 1 & 0 & 1 \\ 
\hline
$\chi^0$ & 1 & 0 & 1 \\ 
$(\rho^+,\rho^0,\rho^-)$ & 3 & 0 & 1 \\ 
\hline
\end{tabular}
\caption{Seesaw dark matter with heavy lepton triplet anchors.}
\end{table}

The model of Table 2 replaces $N_R,N'_R$ with $\Sigma_R,\Sigma'_R$.  
Neutrinos obtain small masses now in the Type III seesaw mechansim 
through the Yukawa terms
\begin{equation}
f_\Sigma (\overline{\Sigma^+_R} \nu_L \phi^+ + \overline{\Sigma^0_R} (\nu_L \phi^0 + l_L \phi^+)/\sqrt{2} + \overline{\Sigma^-_R} l_L \phi^0).
\end{equation}
In the dark sector, the scalar triplet $\rho$ is added to allow the 
Yukawa terms
\begin{equation}
f_\rho \overline{S_L} ({\Sigma'}^+_R \rho^- - {\Sigma'}^0_R \rho^0 + 
{\Sigma'}^-_R \rho^+).
\end{equation} 
Again, for all three neutrinos and $S_L$ to acquire seesaw masses, there 
should be three $\Sigma_R$ copies and one $\Sigma'_R$.  A hybrid variation 
of the two models is clearly also possible with some $N_R,N'_R$ and some 
$\Sigma_R,\Sigma'_R$ adding up to four copies.

\noindent \underline{\it Anomaly Cancellation}~:~ 
Since $S_L$ is the only fermion transforming under $U(1)_D$, the proposed 
dark gauge theory is anomalous.  However, it is crucial that $S_L$ does not 
have a partner with opposite dark charge, or else it would not be massless 
to begin with, which is the first requirement for the seesaw mechanism. 
To allow for this special condition, the following particle content under 
dark $U(1)_D$ may be considered.  Let there be one copy of a singlet with 
one unit of dark charge, i.e. $S_L$, two copies with two units, three 
copies with minus three units, and one copy of four units.  Then~\cite{m21}
\begin{equation}
1(1) + 2(2) + 3(-3) + 1(4) = 0; ~~~ 1(1) + 2(8) + 3(-27) + 1(64) = 0.
\end{equation}
Thus anomaly cancellation is achieved.  Given the charges $q_i$ of the 7 
left-handed chiral fermions, and the unit charge of the scalar $\chi$, the 
only possible fermion mass terms correspond to $q_i+q_j=$ 1 or $-1$. Hence 
only the combinations $2+(-3)=-1$ and $4+(-3)=1$ are allowed.  This means 
that there is a $3 \times 3$ mass matrix linking the 3 fermions of charge 
$(-3)$ to the 2 fermions of charge (2) and the one of charge (4), resulting 
in 3 Dirac fermions.  $S_L$ of charge (1) is left by itself.  For it to have 
a Majorana mass, a Higgs scalar of charge (2) would be needed.  With only 
$\chi$ of charge (1), $S_L$ also cannot couple to the other fermions: (2) 
would require (3), ($-3$) would require (2), and (4) would require (5).  
Hence $S_L$ is massless, which is exactly what is desired for this proposal. 
The situation is analogous to the neutrino case, where the absence of a 
Higgs triplet $(\xi^{++},\xi^+,\xi^0)$ keeps the neutrino massless. 

In this scenario, the lightest dark Dirac fermion is also stable.  This 
accidental symmetry is analogous to that of baryon number and lepton number 
in the SM and is due to the chosen particle content under the dark $U(1)_D$ 
gauge symmetry.  For the purpose of this study, it is assumed that the 
reheat temperature of the Universe is well above the SM Higgs boson mass but 
not high enough to produce these dark fermions.  More discussion on these 
possible dark matter candidates will be presented later.

\noindent \underline{\it Two Higgs Potentials}~:~ 
In the model of Table 1, the Higgs sector consists of the SM doublet $\Phi$ 
and the dark singlet $\chi$.  Its scalar potential is simply given by
\begin{equation}
V_1 = -\mu_0^2 \Phi^\dagger \Phi - \mu_1^2 |\chi|^2 + {1 \over 2} \lambda_0 
(\Phi^\dagger \Phi)^2 + {1 \over 2} \lambda_1 |\chi|^4 + \lambda_{01} 
(\Phi^\dagger \Phi)|\chi|^2.
\end{equation}
As $\phi^0$ and $\chi^0$ acquire nonzero vacuum expectation values 
$v_0$ and $v_1$, the SM gauge symmetry breaks to electromagnetic $U(1)$ 
and the dark $U(1)_D$ symmetry is completely broken.  The minimum of $V_1$ 
is given by
\begin{equation}
\pmatrix{v_0^2 \cr v_1^2} = {1 \over \lambda_0 \lambda_1 - \lambda_{01}^2} 
\pmatrix{\lambda_1 & -\lambda_{01} \cr -\lambda_{01} & \lambda_0} 
\pmatrix{\mu_0^2 \cr \mu_1^2}.
\end{equation}
The $2 \times 2$ mass-squared matrix spanning $\sqrt{2} Re (\phi^0)$ and 
$\sqrt{2} Re (\chi^0)$ is then
\begin{equation}
{\cal M}^2_{\phi \chi} = \pmatrix{2\lambda_0 v_0^2 & 2\lambda_{01} v_0 v_1 \cr 
2\lambda_{01} v_0 v_1 & 2\lambda_1 v_1^2}.
\end{equation}

In the model of Table 2, a complex scalar triplet 
$\rho=(\rho^+,\rho^0,\rho^-)$ is added so that
\begin{equation}
V_2 = V_1 + m_2^2 \rho^\dagger \rho + {1 \over 2} \lambda_2 
(\rho^\dagger \rho)^2 + \lambda_{02} (\Phi^\dagger \Phi)(\rho^\dagger \rho) 
+ \lambda_{12} |\chi|^2 (\rho^\dagger \rho) + [\lambda_{012} 
\chi^\dagger \Phi^\dagger (\vec{\sigma} \cdot \vec{\rho}) \Phi + H.c.].
\end{equation}
Assuming that $m_2^2 >> \mu^2_{0,1}$, the vacuum expectation value 
$\langle \rho^0 \rangle = v_2$ is small~\cite{m01}, i.e.
\begin{equation}
v_2 \simeq {\lambda_{012} v_1 v_0^2 \over m_2^2},
\end{equation}
whereas $v_{0,1}$ remain as given by Eq.~(7) to a very good approximation. 
The $3 \times 3$ mass-squared matrix spanning $\sqrt{2} Re (\phi^0)$,  
$\sqrt{2} Re (\chi^0)$, and $\sqrt{2} Re (\rho^0)$ becomes
\begin{equation}
{\cal M}^2_{\phi \chi \rho} = \pmatrix{2\lambda_0 v_0^2 & 
2\lambda_{01} v_0 v_1 & -2\lambda_{012} v_0 v_1 \cr 2\lambda_{01} v_0 v_1 & 
2\lambda_1 v_1^2 & -\lambda_{012} v_0^2 \cr -2\lambda_{012} v_0 v_1 & 
-\lambda_{012} v_0^2 & m_2^2}.
\end{equation}
 
\noindent \underline{\it Three Key Mixings}~:~ 
In the model of Table 1, the observed scalar boson (call it $h$) at 125 GeV 
is a mixture of the SM Higgs boson and its $U(1)_D$ counterpart according 
to Eq.~(8).  Assuming the latter to be much heavier, this mixing is small, 
i.e. 
\begin{equation}
\theta_{\phi \chi} \simeq {\lambda_{01} v_0 \over \lambda_1 v_1}.
\end{equation}
Another mixing is that of $S_L$ with $N'_R$ coming from Eq.~(2), i.e.
\begin{equation}
\theta_{SN'} = {f_\chi v_1 \over m_{N'}}.
\end{equation}
This is the analog of the $\nu-N$ mixing in the neutrino sector.  The seesaw 
mass of $S$ is
\begin{equation}
m_S = {(f_\chi v_1)^2 \over m_{N'}}.
\end{equation}
The third mixing comes from the assumed small $N N'$ mass terms.  Consider 
for simplicity the $4 \times 4$ mass matrix spanning $(\nu,S,N,N')$.  It 
is of the form
\begin{equation}
{\cal M} = \pmatrix{0 & 0 & f_N v_0 & 0 \cr 0 & 0 & 0 & f_\chi v_1 \cr f_N v_0 & 0 & m_N & \epsilon \cr 0 & f_\chi v_1 & \epsilon & m_{N'}}.
\end{equation}
The reduced seesaw $2 \times 2$ mass matrix is then
\begin{equation}
{\cal M}_{\nu S} = {1 \over m_N m_{N'}-\epsilon^2} \pmatrix{ 
-m_{N'} f_N^2 v_0^2 & f_N v_0 f_\chi v_1 \epsilon \cr  
f_N v_0 f_\chi v_1 \epsilon & -m_N f_\chi^2 v_1^2}.
\end{equation}
Assuming that $m_S >> m_\nu$, the $\nu-S$ mixing is then 
\begin{equation}
\theta_{\nu S} = {f_N v_0 \epsilon \over m_N f_\chi v_1},
\end{equation}
which is indeed very much suppressed. This mixing shows that $S$ may be 
considered a sterile neutrino, but its mixing with the active neutrinos is 
naturally very small.

In the model of Table 2, $h$ mixes with $\rho^0$ according to 
\begin{equation}
\theta_{\phi \rho} \simeq {-2\lambda_{012} v_0 v_1 \over m_2^2} \simeq 
{-2 v_2 \over v_0}.
\end{equation}
Another mixing is that of $S_L$ with ${\Sigma'}^0_R$ coming from Eq.~(4), 
i.e.
\begin{equation}
\theta_{S \Sigma'} = {-f_\rho v_2 \over m_{\Sigma'}}.
\end{equation}
This is the analog of the $\nu-\Sigma^0$ mixing in the neutrino sector.  
The seesaw mass of $S$ is
\begin{equation}
m_S = {(f_\rho v_2)^2 \over m_{\Sigma'}}.
\end{equation}
Now $v_2$ contributes to the $W$ mass shift as measured by 
$CDF$~\cite{cdf22}, i.e.
\begin{equation}
M_W = 80.4335 \pm 0.0094~{\rm GeV}.
\end{equation}
The central value here crresponds to $v_2 \simeq 3.68$ GeV from the 
analysis of Ref.~\cite{ps23}. 
 
The third mixing is of course the $\Sigma \Sigma'$ analog of $N N'$ 
as shown above.

\noindent \underline{\it Rare Higgs Decay to Dark Matter}~:~
In the model of Table 1, $h$ decays to $SS + \bar{S} \bar{S}$ through 
$\theta_{\phi \chi}$ and $\theta_{SN'}$ as shown in Fig.~1.

\begin{figure}[htb]
 \vspace*{-7cm}
 \hspace*{-3cm}
 \includegraphics[scale=1.0]{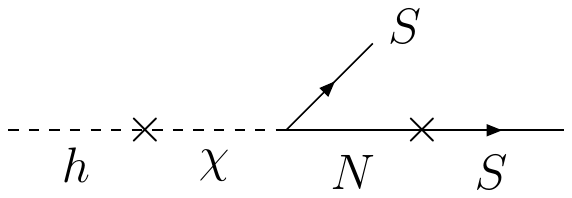}
 \vspace*{-22.0cm}
 \caption{Decay of $h$ to $SS$.}
 \end{figure}

\noindent The effective coupling is
\begin{equation}
f_h = f_\chi \theta_{\phi \chi} \theta_{SN'} = {2 \lambda_{01} v_0 m_S 
\over m^2_\chi}.
\end{equation}
The decay rate of $h \to SS + \bar{S} \bar{S}$ is~\cite{mr21}
\begin{equation}
\Gamma_h = {f_h^2 m_h \over 8 \pi} \sqrt{1-4r^2} (1-2r^2),
\end{equation}
where $r = m_S/m_h$.  Now $S$ is light and a candidate for long-lived 
dark matter.  The correct relic abundance is obtained~\cite{ac13} if 
\begin{equation}
f_h \sim 10^{-12} r^{-1/2}.
\end{equation}  
Hence 
\begin{equation}
m_S = (16~{\rm keV})\left( {m_\chi \over {\rm 5~TeV}} \right)^{4/3} 
\left( {0.005 \over \lambda_{01}} \right)^{2/3}.
\end{equation}

In the model of Table 2, the decay of $h$ to $SS$ also proceeds with 
$\chi$ replaced with $\rho^0$ and $N'$ replaced with ${\Sigma'}^0$ in Fig.~1. 
Hence
\begin{equation}
f_h = f_\rho \theta_{\phi \rho} \theta_{S \Sigma'} = {2 m_S \over v_0}.
\end{equation}
Combining this with Eq.~(24), $m_S = 1$ keV is uniquely determined.

\noindent \underline{\it Long-Lived Dark Matter}~:~
Since $S$ is a singlet fermion, it may be considered a sterile neutrino, 
but its mixing with the active neutrinos, i.e. $\theta_{\nu S}$ of Eq.~(17), 
is suppressed by $N N'$ or $\Sigma \Sigma'$ mixing which breaks the 
separate conservation of lepton parity and dark parity as already pointed out. 
It may therefore decay radiatively~\cite{pw82} into a neutrino and a photon.
Its lifetime is then given by
\begin{equation}
{\tau_\gamma \over \tau_U} = 4.2 \times 10^3 ~\theta_{\nu S}^{-2} \left( 
{1~{\rm keV} \over m_S} \right)^5,
\end{equation}
where $\tau_U = 4.35 \times 10^{17}$s is the age of the Universe. 
However, for decaying dark matter not to conflict with the observations of 
the Cosmic Microwave Background (CMB), $\tau_\gamma > 10^{25}$s is 
required~\cite{sw17}.  Hence
\begin{equation}
\theta_{\nu S} \left( {m_S \over 1~{\rm keV}} \right)^{5/2} < 0.014.
\end{equation}
This shows that $m_S$ should not be much greater than 10 keV or 
$\theta_{\nu S}$ would have to be chosen to be extremely small.

\newpage
\noindent \underline{\it Heavy Fermion Dark Sector}~:~
Back to the 3 Dirac fermions with $U(1)_D$ charges $(-3,-3,-3)$ paired 
with $(2,2,4)$ through the Higgs scalar $\chi$.  They possess an accidental 
unbroken global U(1) symmetry at the Lagrangian level, and the lightest 
(call it $\psi$) could be a dark-matter candidate.  However the 
gauge-invariant dimension-five term
\begin{equation}
{\cal L}_5 = {\psi S \chi \chi \over 2 \Lambda} + H.c.
\end{equation}
breaks this global symmetry for $\Lambda$ at the Planck scale~\cite{ck21} 
and allows the decay $\psi \to Sh$ with 
$f_\psi = v_1 \theta_{\phi \chi}/\Lambda_{Pl}$. 
The decay rate is $\Gamma_\psi = f_\psi^2 m_\psi/8\pi$.
Let $\lambda_{01}/\lambda_1=0.1$, $\Lambda_{Pl} = 2.4 \times 10^{18}$ GeV, 
then $m_\psi > 1$ TeV implies $\tau_\psi < 3 \times 10^8$s.

Hence quantum gravity rules out $\psi$ as a dark matter candidate.  This 
argument has the same origin as that against massless fermions, and applies 
to all global and discrete dark symmetries and tends to favor 
\underline{light dark matter}, which is the subject of this paper.

\noindent \underline{\it Concluding Remarks}~:~
Dark matter is postulated as akin to neutrinos.  Both are massless at the 
Lagrangian level, having no invariant masses due to gauge symmetry. 
Whereas neutrinos transform as components of SM doublets, dark matter 
is a fermion singlet $S$ transforming under a postulated dark $U(1)_D$ gauge 
symmetry, which is spontaneously broken by a scalar singlet.  Both obtain 
small seesaw masses from either heavy neutrino singlets or heavy lepton 
triplets.  In the latter case, an additional scalar triplet is needed, 
whose small vacuum expectation value allows the $W$ mass to be shifted 
upward relative to that of the SM, which is a possible explanation of 
the recent $CDF$ precision measurement.  The resulting dark matter 
candidates in either case are possible freeze-in products of the decay 
of the observed Higgs boson of 125 GeV, provided the reheat temperature 
of the Universe is well above this mass but well below the breaking 
scale of $U(1)_D$.  

In the case of heavy singlet seesaw anchors, $m_S = 16$ keV is considered 
as an example. In the case of heavy triplet seesaw anchors, $m_S = 1$ keV 
is uniquely determined. Both may be considered singlet neutrinos with 
very small $\theta_{\nu S}$ mixing and decay to a neutrino and a photon.
Unlike the canonical scenario with sterile neutrinos as warm dark matter, 
the production mechanism of $S$ through Higgs decay does not depend 
on $\theta_{\nu S}$ whereas its decay rate does, thereby evading the strong 
constraints in the case of sterile neutrinos where both depend on 
$\theta_{\nu S}$.  Although the low-energy phenomenology is indistinguishable 
from just adding a light sterile neutrino to the SM, the current proposal 
offers a theoretical understanding to why the sterile neutrino is light, 
and the hope that at energies beyond a few TeV, the dark gauge sector may 
be observed.

\noindent \underline{\it Acknowledgement}~:~
This work was supported in part by the U.~S.~Department of Energy Grant 
No. DE-SC0008541.  

\baselineskip 18pt
\bibliographystyle{unsrt}

\end{document}